\documentclass{article}

\usepackage[utf8]{inputenc}
\usepackage[T1]{fontenc}
\usepackage{microtype}
\usepackage{xcolor}
\usepackage{newspaper}
\usepackage{tikz}

\usetikzlibrary{shapes.arrows, fadings}

\date{April 1, 2021}
\currentvolume{1}
\currentissue{1}

\SetPaperName{The Swapland}

\SetHeaderName{The Swapland}

\SetPaperLocation{Cambridge, MA}
\SetPaperSlogan{ \em ``The arXiv digest you wish you had every day''}
\SetPaperPrice{Zero Dollars}

\usepackage{graphicx}
\usepackage{multicol}

\usepackage{picinpar}

\usepackage{newspaper-mod}
\renewcommand{\headlinestyle}{\scshape\Large\lsstyle}

\usepackage{lipsum}

\begin{document}
\maketitle

{\centering \Huge\bfseries\boldmath{$\mathrm{R_K}$ opens; All eyes on g$-$2 IPA next week}}

\begin{multicols}{2}
\paragraph{Rumors fly} \hspace{-0.5em}as the Fermi National Accelerator Laboratory filed papers announcing a potentially multisigma Initial Public Anomaly for the muon g$-$2. This is a milestone for the long-standing tension, who has captured promising investment since its inception at Brookhaven National Lab. As of press time, neither a measurement nor SM consistency have been specified.

The anticipation follows the successful opening of the LHCb B-decay anomalies last week, debuting at a respectable 3$\sigma$, reduced margins, and high initial investment from the community. The flavor anomaly, first established at the SLAC Accelerator Lab, is notable in having maintained its high significance through the years where many others have faded away...

\noindent\includegraphics[width=\linewidth]{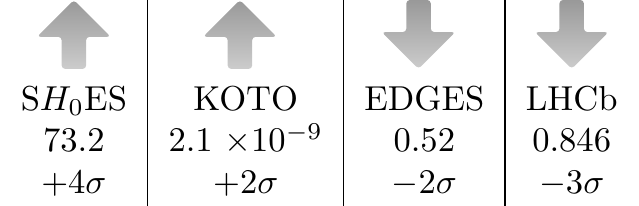}

\end{multicols}

\closearticle

\begin{multicols}{2}

{\centering \huge \textbf{Archaeologists Unearth \& Translate Ancient Codes}}

\vspace{2mm}

\paragraph{Archaeologists in New England have discovered} \hspace{-0.5em}what they believe to be a copy of ancient SUSY code dating back to at least 1998. The relic, extracted painstakingly from an old hard drive in some unnamed theorist's office last Thursday night, is written in a particularly archaic branch of Fortran.  The code is conjectured to have computed something related to a particular particle model, though how and why is for now entirely beyond mortal ken. Translation is underway but prospects are not optimistic, as linguistic traits point to pre-SLHA orthography, and preliminary efforts hint at an extraordinary amount of obfuscating syntax even for its historical era. As of press time, researchers are unsure why they're doing this at all...

\closearticle 

{\centering \huge \textbf{New Proposal Aims to Bring Muons Together}}

\vspace{2mm}

{\bf A new building proposal with multi-institutional support is gaining headway, sources report, and at its heart a goal to bring billions of muons together.}

The plan, which would demand a state-of-the-art convention center geared exclusively towards muon-led events, claims that such a locale would guarantee a veritable explosion of cultural- and virtual-particle exchange for these historically ``unwanted" particles. Although a location for this event center has yet to be specified, the design promises a precision-driven and high-energy environment that centers the well-being of its inhabitants -- the core concern to most contributors is the stability of the incoming patrons. Proponents argue that this new establishment will serve to bring muons together rather than drive them apart, which has historically been the overwhelming occurrence in electron- and proton-dominated spaces... 

\closearticle

{\Large \scshape Opinions \& Local News}

\vspace{2mm}

{\centering \large \textbf{If I have one more Zoom meeting today I am going to stab my eyes out}}

\vspace{2mm}

{\centering \large \textbf{Area physicist deliberates whether to attend in-person workshop at Colorado ski resort}}

\vspace{2mm}

{\centering \large \textbf{Review: Latest seminar lacks emotional throughline, relatable characters, compelling action}}

\closearticle

{\Large \scshape Sports}

\vspace{2mm}

{\centering \large  \bf Olympic Speaker Breaks Personal 14h Record}

In a startling feat of human athleticism, the former Olympic gold medalist has broken his yet unchallenged personal record for seminar speaking time, leaving onlookers stunned...

\closearticle

{\centering \large  \bf Contact the Editors}

Please send in your letters to the following address:

\begin{center} 
{\it  Harvard Dept. of Physics}

{\it 17 Oxford St, Cambridge, MA, 02138}
\end{center}

We welcome all community input. The Board of Editors are: P. Agrawal, H. Bagherian, C. Cesarotti, N. DePorzio, Q. Lu, J.~B. Mu\~noz, A. Parikh, M. Reece, and W. L. Xu.

\end{multicols}

\newpage 

\begin{center}
{\Huge \scshape \textbf{Classified AdS}}
\end{center}

\closearticle

\begin{multicols}{3}

\headline{Free and For Sale}

\paragraph{Box of Dark Matter models} FREE!! accidentally acquired too many of them. Fair condition, may require some tuning. Must take all of them at once, as some are broken or missing UV completions. 

\paragraph{Supersymmetry} Beautiful and mint condition, but we just never got the chance to use it. Looking for someone good with modeling to turn it into something more practical. Originally bought for 1 LHC, price negotiable. No strings attached.

\paragraph{Vintage Technicolor theories} A set of old-fashioned gauge theories to bring whimsy to your home. Flavor-changing for a refined palette. Warning: not compatible with most editions of LEP.

\paragraph{Several grad students} Sold individually or as a set. Cheap, please hire us.

\paragraph{GUTs} Buyer beware, don’t bring your precious protons too close.

\paragraph{Machine Learning} Take a GANder at these neural nets. Ready to be used out of the box on your favorite physics problem, like telling apart pictures of cats and dogs!! Free courtesy of our computer science friends.

\paragraph{Some pots} Moderately to severely cracked. Great for recreational purposes! Caution is advised, as prolonged use may result in errant discoveries and long, angry emails.

\paragraph{Gravity} FREE! Modified, does not work any more. Weight: 1 stone with some extra scalars.  Ideal for someone with a lot of functions. May be safe.

\paragraph{For Sale:} effective theories, never completed.

\closearticle

\headline{Missed Connections}

\paragraph{}Me: healthy infrared theory, in search of a completion. You: beautiful ultraviolet theory. You think we have incompatible 't Hooft anomalies, but what if I dress in the right TQFT?

\paragraph{}Me: A Magnetic Monopole. You: A scientist. That Valentine's night, I passed through your superconducting ring. I am sorry if I left you in flux.

\paragraph{}I observed Wigner observing you, but I don't know what state you're in. Message me, before I collapse.

\paragraph{}Me: A phenomenologist. You: An astronomer. I saw some data in a plot you flashed in a talk and I think I can explain your 2-$\sigma$ anomaly with my 5-parameter BSM model. PRL and chill?

\closearticle

\headline{Services}

\paragraph{Tooth Fairy Rentals} Get your wildest model-building dreams fulfilled, for the low price of your self-respect as a scientist. Financing options available, no payment due before tenure. Wish-fulfillment may not extend to reality. Limit one per customer.

\paragraph{Health Trainer} Expert advice from a former renowned model. A traditional fitness routine focusing on running to high energies along with a regimen of proton-biotics. Guaranteed to improve your GUT health.    

\paragraph{Dyson Vacuum Service} We clean while you wait. We use the most sustainable energy harnessed from a nearby star. No blades.

\paragraph{Street Sweeper} For hire, will sweep through your entire space and locate all anomalies. As a courtesy we include a look at all street bins, roadside pileup, and elsewhere. Note: only willing to work under lampposts. 

\paragraph{Tuner} Trade in your uncontrolled moduli for brand new, finely tuned scalars! To get more information call 1-800-SLOW-ROLL.

\closearticle

\headline{Career Opportunities}


\paragraph{Data Scientist} Sell your soul, but not completely. Better than finance, and since you're already applying Machine Learning to your physics, you can even continue using something you learned for research!

\paragraph{Software Engineer} What the other guy said, but more money.

\paragraph{Babysitter} Needs to be able to take care of multiple baby universes. Warning: any information you give them will be lost.


\paragraph{Quantum Person} If you Quantum, we want you. Ultra Quantum preferred.

\paragraph{Operator} At Reeh-Schlieder enterprises, our motto is ``think globally, act locally.'' Help us create a headquarters on the moon, without leaving the comfort of your home office.

\paragraph{Wordsmith} Obsessed with wordplay? Join the premier PR department tasked with determining the future of experimental acronyms in our field and help us GAUGE:
Generate
Acronyms for
Upcoming
Galaxy-probing
Experiments

\closearticle

\pagebreak

\headline{Lost and Found}

\paragraph{Lost:} Nonsupersymmetric AdS vacuum. Spacious vacuum, carefully constructed and orbifolded, last spotted disappearing into a bubble of nothing. Bag of gold as reward.

\paragraph{Found:} Evidence of point sources in the galactic center. Flashing quickly (and periodically), will probably become diffuse by next week.

\paragraph{Found:} 2.5$\sigma$ anomaly. Adorable little bump found abandoned between bins of X-ray data. Comes from a bad background, but has the potential to become significant if properly cleaned up! In need of a good home and a lot of attention.
 
\paragraph{Found:} Flying rock, cigar shaped. Careful, may be icy. Please Contact Dr.~Ellie Arroway for further details.

\paragraph{Lost, \emph{never} Found:} Factors of 2, $\pi$, i and minus signs. 
 
\closearticle

\headline{Public Notices}

\paragraph{Call for Information} on Black Holes. Any photographs, recordings, legal documentation and knowledge of past entanglements requested at high prices.

\noindent\includegraphics[width=\linewidth]{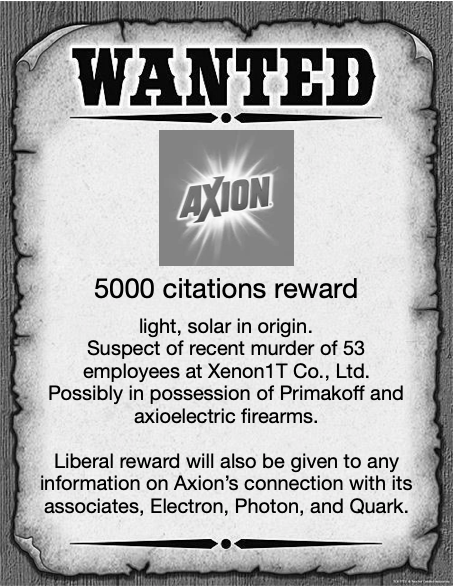}





\end{multicols}

\closearticle

\begin{center}
{\Huge \scshape \textbf{Lifestyle \& Entertainment}}
\end{center}

\closearticle

\begin{multicols}{3}

\byline{Advice Column}{Dr. Snowmass}

Hello Colleagues, thank you for all of your letters in the past year. Let's see what we have today...

\paragraph{Dear Dr. Snowmass,} Just wanted to say hi. Good to meet everyone!

\hfill \textit{-- Anon}

\textit{Dear Anon,} Likewise! \hfill \textit{-- Dr. S.}

\paragraph{Dear Dr. Snowmass,} I would like to draw your attention to a particular particle model that I believe has been grievously overlooked by the community. I refer you to Solo Author Papers [1-16] attached below, which I believe should constitute major sections in the White Papers for every Frontier imaginable. \hfill \textit{-- Deserves to be cited}

\textit{Dear Deserves,} Thank you for your submission; your letter was the 57th this week trying to use Snowmass as a platform for self-promotion.  While I'm sure your work is groundbreaking this is getting annoying, please find attached my formal Letter of Disinterest. \hfill \textit{-- Dr. S.}

\paragraph{Dear Dr. Snowmass,} Hello all, excited to be here! \hfill \textit{-- Anon 2}

\textit{Dear Anon 2:} Welcome! \hfill \textit{-- Dr. S.}

\paragraph{Dear Dr. Snowmass,} I received tenure 17 years ago but I firmly believe I still have all the scientific vivacity of my youth! I have fresh ideas, am interested in new technologies, and most importantly, still feel like I have no idea what I'm doing. Would it be OK to brand myself as an early career scientist? \hfill \textit{-- Junior faculty at heart}

\textit{Dear Junior:} Judging by the fact that you wrote to a newspaper advice column, I'm afraid we have differing judgments on the topic of your youth. \hfill \textit{-- Dr. S.}

\paragraph{Dear Dr. Snowmass,} Hi!!
\hfill \textit{-- Anon 3}

\textit{Dear Anon 3,} :waves: \hfill \textit{-- Dr. S.}

\paragraph{Dear Dr. Snowmass,} I recently stumbled upon the secret to cold fusion. What would be the most effective way to communicate this discovery to the community at large?  \hfill \textit{ -- Hidden Genius}

\textit{Dear Hidden,} While Snowmass is not the most appropriate avenue to advertise original research, I can however give some very specific recommendations on what you should \textit{not} do.  \hfill \textit{-- Dr. S.}

\paragraph{Dear Dr. Snowmass:}  If ER = EPR, doesn’t P = 1?
\hfill \textit{-- Algebraist}

\textit{Dear Algebraist,} ER = 0 happens to work as well. The Clay Mathematics Institute might also be interested in your thoughts on P = NP. \hfill \textit{-- Dr. S.}

\paragraph{Dear Dr. Snowmass:} I am Groot.

\hfill \textit{-- Groot}

\textit{Dear Groot,} For the sake of scientific transparency, please use your professional name in future correspondence. \hfill \textit{-- Dr. S.}

\closearticle 

\end{multicols}

\newpage

\begin{multicols}{3}

\headline{Recipes for the Week}

\paragraph{Sourdough Starter} This starter will let you bake a Universe with a deep and cultivated sense of dread.
In a small comoving volume, mix together 5 parts of dark matter per each one of baryons, and stir well until only adiabatic fluctuations remain.
Keep the mixture at room (photon) temperature, which will allow it to absorb entropy as time progresses, deepening the flavor.
By now you may need to move your starter to a larger container, or simply expand the Universe itself.
Keep the mixture in a Hubble jar with a loosely fit lid, and remember to feed with vacuum energy every once in a while.
Some chefs keep their starters going on for millions, or even billions of years, and they claim to produce anthropically delicious results.

\paragraph{Linguini con Sugo di Pomodoro}
Before selecting a recipe, one should thoughtfully peruse the arXiv for fresh produce and meaningful ideas. Food is so much more delicious when made by someone who has taken the time to fully consider the nuances required to complete the dish. Cook with only ingredients you have examined and know to be fresh and authentic. 

Begin by sculpting a mound of flour, sound enough to contain your eggs and oil without immediately devolving into a sticky, unstructured mess. Carefully stir to combine, making sure to leave no uncombined flour or unrelated tangents. To develop the gluten, knead the pasta for several minutes or pages until the dough is easily identifiable as such. Let rest for ten minutes. 

Begin to roll your dough out carefully, making sure to not jump dramatically in thickness or in logical reasoning. A continuous deformation will lead to tastier results with fewer inconsistencies. Once at desired thickness and length, cut into strips and include an Acknowledgement section. Bring a large pot of water to boil. 

At the same time, heat the diced tomato in a large skillet with any other vegetables that pair nicely. Do not include any that were out of season and rotten. Cook down to a rich sauce and season. 

Boil the linguini until slightly before al dente, not to the point of being overdone and unoriginal. Move the pasta to the skillet with the sauce, along with a few spoonfuls of pasta water, to finish cooking. Serve hot, with fresh parmigiano and appendices explaining any convoluted, hidden equations. 

\paragraph{Linguini con Sugo di Pomodoro (\textit{Ambulance Chaser Version})}
\textit{AKA Pasta with Red Sauce. }
Dump whatever noodles fit into the last clean bowl in your kitchen. Fill the bowl with enough water to cover, then chuck in the microwave until boiling. Haphazardly drain water, and replace with mass-produced chunky red sauce out of a plastic jar. Stir, serve scalding hot but simultaneously room temperature. Ready as soon as you need it. 

\paragraph{ Cocktail: The \emph{Thesis Writer}} \phantom{a} \newline

\textit{Ingredients.}  Bourbon in a coffee mug

\textit{Notes.} Look, we get it.

\paragraph{ Cocktail: The \emph{Job Application}} \phantom{a} \newline

\textit{Ingredients.}  3 oz of publications. 
1 job talk, finely aged. 
 Dash of rejection bitters. 
1 shortlist of better candidates, steeped overnight in impostor syndrome.
Compulsory rumor-mill refreshing.
3-5 letters of recommendation.

\textit{Notes.} Shake well, garnish with existential dread.

\paragraph{ Cocktail: The \emph{Simulation}} \phantom{a} \newline

\textit{Ingredients.} 
Dark matter, diced as finely as budget allows.
Baryons, lightly carbonated (optional).
Hydrodynamic feedback, to personal taste.
A dash of magnetic fields.

\textit{Notes.} Combine ingredients and allow to sit for several million CPU hours.
Small clumps may appear, this is normal. This wonderfully simple---and deceptively expensive---recipe is guaranteed to please your illustrious (TNG) dinner guests. 
The dark matter is traditionally incorporated cold, but try it warm, hot, or fuzzy for diversity's sake!
The most daring palates venture to modify gravity itself.
Feedback, with its FIRE-y aftertaste, is the most delicate part of this recipe; too much and it could easily overpower, too little and one feels that something is missing. The art is adding in just enough to keep it from the cusp of bitterness so that the core flavors can be experienced.  

\closearticle

\headline{Horoscope}
\begin{center}
    \includegraphics[width=0.3\linewidth]{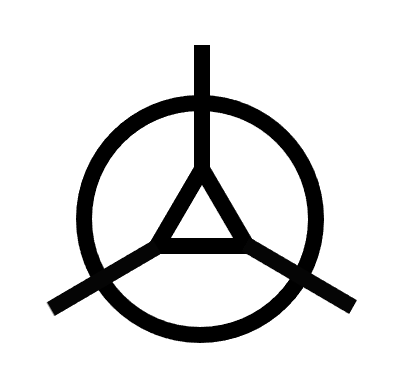} \\
    \textbf{Arecibus} \\ 
    \textbf{August 10 - November 6}
\end{center}
\paragraph{Arecibus} injuries and illness are anticipated. You have achieved great success in life, but you now feel as if your world is caving in around you. While it seems like you are hanging on by a thread, you have weathered many storms and often prevail when you set your mind to it. You should know that transformation and new beginnings are highly favorable to Arecibus at this time. Consider having children, or puttering your resources towards new causes. Your energy is particularly aligned with the far side of the Moon this cycle. Place your meditations there to identify new projects. 

\closearticle
\end{multicols}

\newpage

\headline{Crossword}

\begin{center}
\includegraphics[width=0.75\textwidth]{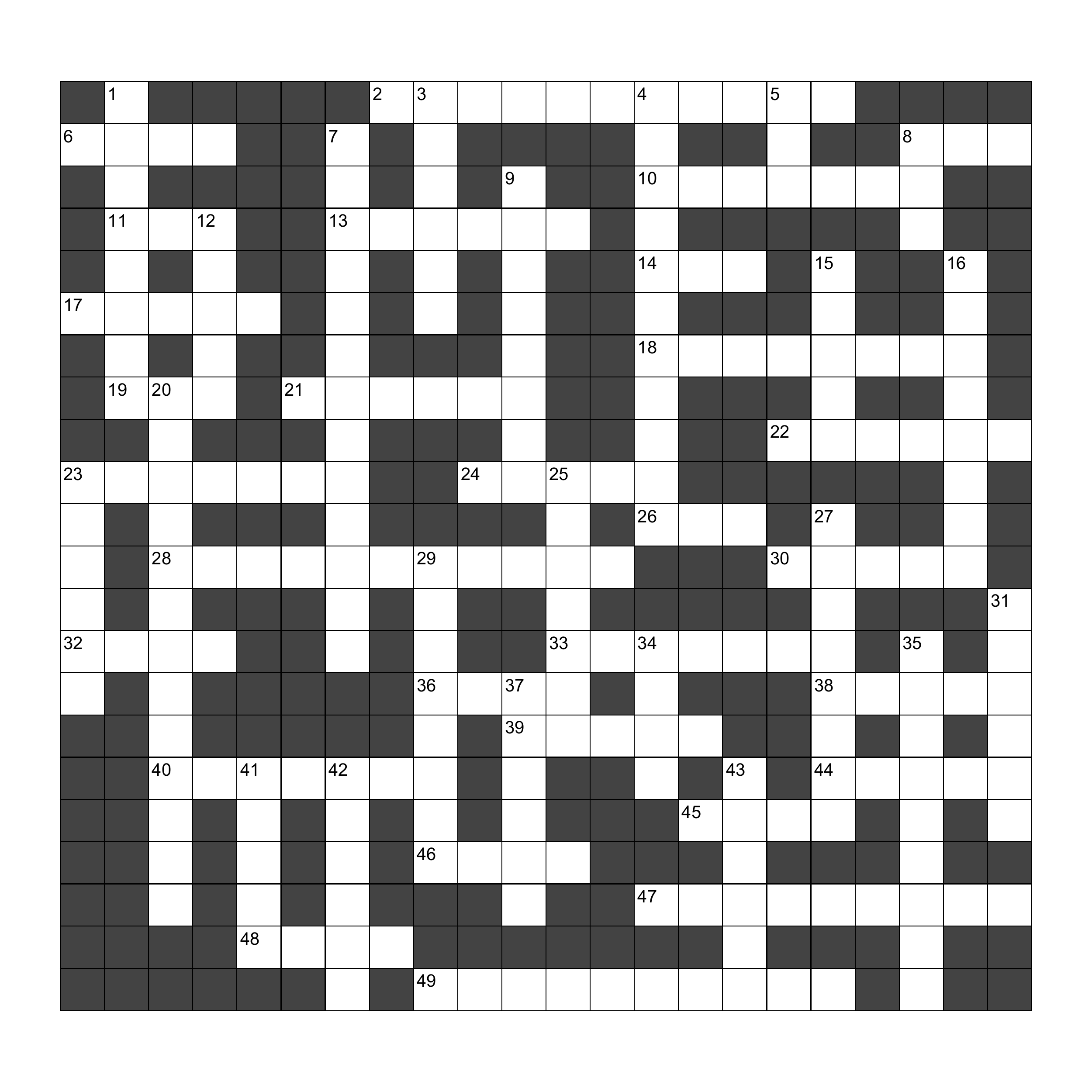}
\end{center}

\noindent\begin{minipage}[t]{0.55\linewidth}
\vspace{0pt}
\begin{multicols}{2}
\small
{\scshape \large Across}
\begin{enumerate}\itemsep-0.6mm
\item[2] Wurst name for a galactic object 
\item[6] Not dressed 
\item[8] COBE's beef  
\item[10] Ever Given in the Suez
\item[11] Casually greeting a scalar quark 
\item[13] A type of universe \& a type of monster  
\item[14] Quit Dynkin around in simple laces
\item[17] Dimensional analysis
\item[18] Sometimes mixed with 4 Down
\item[19] Emotion-Focused Therapy for low-energy physicists
\item[21] Oracle of Delphi prophesizing new physics 
\item[22] Dirac adopts a big red dog 
\item[23] One step at a time 
\item[24] Trajectory narrowly misses Jamaica 
\item[26] Alternative truth 
\item[28] A Creek Owlet confused about fundamental forces 
\item[30] Unitarily measured (not unit of measurement)
\item[32] Don't shame this defect 
\item[33] No more Fyre-walls 
\item[36] Type of error (abbr) 
\item[38] Australian bear learns machining (sic) 
\item[39] Sphere vacuum
\item[40] Baby frog in a penguin family 
\item[44] Collider one can find in California
\item[45] Hard to balance all the mass on a point 
\item[46] Moved anomalously, like Britain 
\item[47] Diluter of curvature and currency
\item[48] Cooperate in field of mutual interest
\item[49] Dinosaur killer 
\end{enumerate}
\end{multicols}
\end{minipage}
\hfill\vline\hfill
\begin{minipage}[t]{0.42\linewidth}
\vspace{0pt}
\begin{multicols}{2}
\small
{\scshape \large Down}
\begin{enumerate}\itemsep-0.6mm
\item[1] Wanted: Dead and Alive
\item[3] Bead counter/boomer calculator
\item[4] Find completion at the end of a rainbow
\item[5] Confinement needs minding
\item[7] Arcade Fire song
\item[8] Collider experiment, but not facility
\item[9] Studied with ferocity, no, fericity
\item[12] Getting a couch up the stairs with friends
\item[15] To avoid, lest your theories get bogged down
\item[16] Kinematic feature to measure masses
\item[20] Brother skips 2 generations 
\item[23] Three for Muster Mark 
\item[25] Sandra Bullock \& George Clooney space drama
\item[27] You run to it (and to get rid of it) 
\item[29] Virtually divested from fossil fuels
\item[31] 21 micrograms
\item[34] Unknotting quantum corrections
\item[35] 196883 reasons to get drunk cheaply
\item[37] You start with the brane, and the brane is BPS
\item[41] Prophet to no god
\item[42] Close but no cigar
\item[43] Stem of (anag) your Apple
\end{enumerate}
\end{multicols}
\end{minipage}

\newpage

\headline{Comics}
\begin{center}\includegraphics[width=\textwidth]{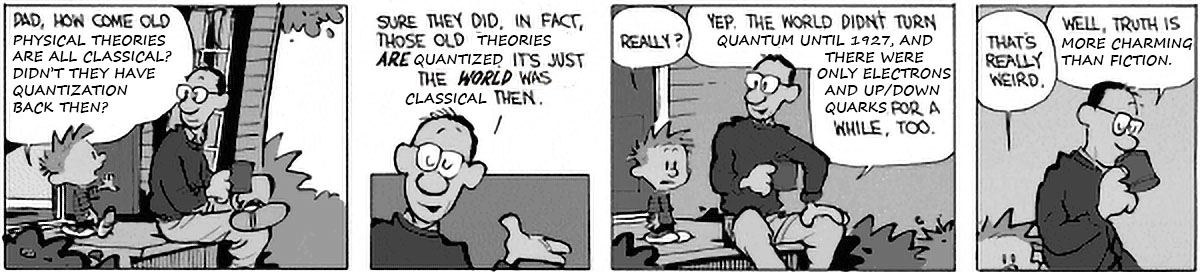}

\vspace{3mm}
\noindent\includegraphics[width=\textwidth]{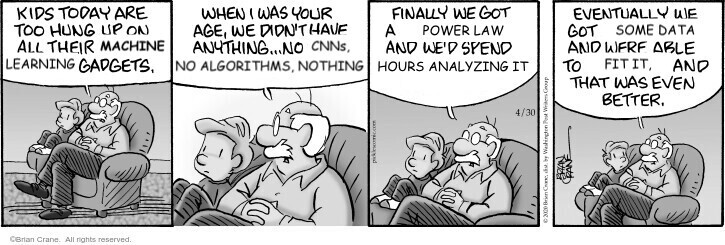}
\end{center}

\closearticle

\headline{Solutions to the Crossword}

\rotatebox[origin=c]{180}{%
\noindent
\begin{minipage}[t]{0.49\linewidth}
\vspace{0pt}
\begin{multicols}{2}
\small
{\scshape \large Across}
\begin{enumerate}\itemsep-0.5mm
\item[2] Gaia sausage 
\item[6] bare
\item[8] CMB
\item[10] trapped
\item[11] sup
\item[13] pocket
\item[14] ADE
\item[17] naive
\item[18] infrared
\item[19] EFT
\item[21] Pythia
\item[22] spinor
\item[23] quantum
\item[24] Regge
\item[26] top
\item[28] electroweak
\item[30] qubit
\item[32] kink 
\item[33] islands
\item[36] stat
\item[38] cwola
\item[39] Dyson
\item[40] tadpole
\item[45] pole
\item[46] left
\item[47] inflation
\item[48] cite
\item[49] dark matter
\end{enumerate}
\end{multicols}
\end{minipage}%
\begin{minipage}[t]{0.49\linewidth}
\vspace{0pt}
\begin{multicols}{2}
\small
{\scshape \large Down}
\begin{enumerate}\itemsep-0.5mm
\item[1] cat state
\item[3] abacus
\item[4] ultraviolet
\item[5] gap
\item[7] supersymmetry
\item[8] CDF
\item[9] jet shape
\item[12] pivot
\item[15] swamp
\item[16] endpoint
\item[20] fraternal twin
\item[23] quarks 
\item[25] gravity
\item[27] GUT scale
\item[29] offshell
\item[31] Planck
\item[34] loop
\item[35] moonshine
\item[37] ADS/CFT
\item[41] Dirac
\item[42] oblate
\item[43] MOSFET
\item[44] linac
\end{enumerate}
\end{multicols}
\end{minipage}%
}

\end{document}